# TWO DIMENSIONAL SPARSE-REGULARIZATION-BASED INSAR IMAGING WITH BACK-PROJECTION EMBEDDING

*Xu Zhan, Xiaoling Zhang, Shunjun Wei, Jun Shi*

School of Information and Communication Engineering
University of Electronic Science and Technology of China,
Chengdu, China 611731

## ABSTRACT

Interferometric Synthetic Aperture Radar (InSAR) Imaging methods are usually based on algorithms of match-filtering type, without considering the scene's characteristic, which causes limited imaging quality. Besides, post-processing steps are inevitable, like image registration, flat-earth phase removing and phase noise filtering. To solve these problems, we propose a new InSAR imaging method. First, to enhance the imaging quality, we propose a new imaging framework base on 2D sparse regularization, where the characteristic of scene is embedded. Second, to avoid the post processing steps, we establish a new forward observation process, where the back-projection imaging method is embedded. Third, a forward and backward iterative solution method is proposed based on proximal gradient descent algorithm. Experiments on simulated and measured data reveal the effectiveness of the proposed method. Compared with the conventional method, higher quality interferogram can be obtained directly from raw echoes without post-processing. Besides, in the under-sampling situation, it's also applicable.

## 1. INTRODUCTION

Interferometric synthetic aperture radar (InSAR) provides a powerful tool for obtaining terrain height information [1]. Currently, the interferogram is obtained based on imaging results of match-filtering (MF) type algorithms, like range-Doppler (RD) algorithm. However, some deficiencies exist: 1) limited imaging quality; 2) full-sampling is required; 3) characteristics of the scene are not into consideration; 4) post-processing is not avoidable, including steps like image registration, conjugate-multiplication, removing the flat-earth phase and phase noise filtering, etc. These deficiencies motivate us to develop a new method, where higher quality interferogram can be obtained directly from echoes without post-processing steps, as illustrated in Fig. 1.

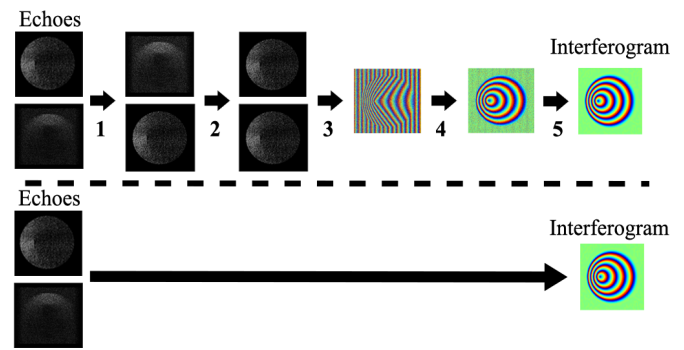

Fig. 1 Conventional method (top) vs the proposed method (bottom). Step 1-5 are match-filtering imaging, registration, conjugate multiplication, flat-earth phase removal and phase noise filtering respectively. Within our method, the high-quality interferogram is obtained directly from raw echoes.

In [2], a target-oriented imaging framework is proposed for automatic recognition task, utilizing merits of sparsity-regularized reconstruction. Inspired by their work, with the high-quality interferogram as orientation, we establish the new InSAR imaging framework based on two-dimensional sparse regularization, where we embed the scene's intrinsic characteristics of 2D slow-varying terrain.

The core issue is to establish the forward observation, which can be approximated by MF type imaging algorithms [3]. We embed the back-projection algorithm (BP) by multiple imaging operations into the forward observation for its multiple merits, like high-precision imaging and parallel calculation capacity. Besides, both flat-earth phase removal and image registration are inherently accomplished in the imaging process [4]. Further, to solve the reconstruction problem, we derive the adjoint operations of forward observation operations, and solve it by forward-backward iterative based on proximal gradient descent [5].

Experiments on simulated data and measured data from an air-borne InSAR system are used to verify the method. Results reveal that in comparison with the conventional method, a higher quality interferogram is directly obtained from raw echoes. Besides, when the raw echoes are under-sampled, our method is still effective.

## 2. 2D SPARSE REGULARIZATION INSAR IAMGING

In this section, we develop the mathematical imaging model of the proposed method.

### 2.1 Conventional imaging method.

The relationship between the scene's scattering coefficients $\mathbf{X}$ and echoes can be expressed as:

$$\mathbf{Y}_m = g(f_m(\mathbf{X} \odot \mathbf{P}_m)) + \mathbf{N}$$
$$\mathbf{Y}_s = g(f_s(\mathbf{X} \odot \mathbf{P}_s)) + \mathbf{N} \quad (1)$$

where $\mathbf{Y}_m$, $f_m(\cdot)$ and $\mathbf{P}_m$ are echo matrix, core forward observation process and topographic phase matrix of the master antenna, while $\mathbf{Y}_s$, $f_s(\cdot)$ and $\mathbf{P}_s$ are those of the slave antenna. $g(\cdot)$ is the sampling operator. $\mathbf{N}$ indicates the noise. $\odot$ denotes Hadamard product. The MF imaging operator can be seen as the least-square-error estimation, which can be expressed as:

$$\mathbf{X} \odot \mathbf{P}_m = \arg\min_{\mathbf{X} \odot \mathbf{P}_m} \frac{1}{2} \|\mathbf{Y}_m - g(f_m(\mathbf{X} \odot \mathbf{P}_m))\|_F^2$$
$$\mathbf{X} \odot \mathbf{P}_s = \arg\min_{\mathbf{X} \odot \mathbf{P}_s} \frac{1}{2} \|\mathbf{Y}_s - g(f_s(\mathbf{X} \odot \mathbf{P}_s))\|_F^2 \quad (2)$$

Then the rough interferogram is formed as:

$$\hat{\mathbf{\Theta}} = (\mathbf{X} \odot \mathbf{P}_m) \odot \operatorname{conj}(h(\mathbf{X} \odot \mathbf{P}_s)) \quad (3)$$

where $h(\cdot)$ is the image registration operator, and $\operatorname{conj}(\cdot)$ is conjugation operator.

Finally, after removing flat-earth phase and phase noise filtering, the fine interferogram is obtained:

$$\mathbf{\Theta} = N(R(\hat{\mathbf{\Theta}})) \quad (4)$$

where $R(\cdot)$ and $N(\cdot)$ indicate flat-earth phase removing operator and noise filtering operator respectively.

### 2.2 2D sparse regularization imaging method.

Seen from (2), the objective contains purely least-square-error constraint. To ensure the stableness of resolution, the sampling scheme needs to obey Nyquist sampling law. Also, there are inevitable sidelobes, clutter and noise. The scene's characteristics are not concerned or enhanced, which results the imaging quality is limited. Thus, to solve this problem, we reform the framework into a 2D sparse regularization one. Combining (1), (3) and (4), the whole forward observation process can be expressed as:

$$\mathbf{Y}_m = f_m\left(g\left(h^{-1}\left((\mathbf{P}_s \odot \mathbf{X}) \odot R^{-1}(\mathbf{\Theta})\right)\right)\right) + \mathbf{N} \quad (5)$$

where $h^{-1}$ and $R^{-1}$ are the adjoint operations of image registration and flat-earth phase removing respectively.

The terrain of imaging scene is normally stationary and homogeneous in local area [6]. This characteristic can be described as 2D sparsity in other domains, like frequency or wavelet domain. To embed this, we solve the interferogram by:

$$\mathbf{\Theta} = \arg\min_{\mathbf{\Theta}} \frac{1}{2} \|\mathbf{Y}_m - A(\mathbf{X}_s \odot R^{-1}(\mathbf{\Theta}))\|_F^2 + \lambda \|\Psi(\mathbf{\Theta})\|_1 \quad (6)$$

where $A(\cdot) = f_m(g(h^{-1}(\cdot)))$ and $\Psi(\cdot)$ is the 2D domain transform operation. $\|\cdot\|_F$ and $\|\cdot\|_1$ are the matrix Frobenius norm and $L_1$ norm respectively. $\lambda$ is the regularization coefficient.

The core issue is to construct the forward observation $A(\cdot)$. We adopt the approximation technique. Specifically, adjoint operators of a MF imaging algorithm procedures are chosen to approximate $A(\cdot)$ that satisfies $I = M^{-1} \approx A$, where $M^{-1}$ indicates the combination of adjoints imaging operators [3]. Specifically, to avoid trivial post-processing steps, the BP is chosen. BP has merits of high precision imaging, parallel calculating capacity, no need to correct range cell migration. What's more, during the imaging process, imaging results can be directly coded at the same grid, and the flat-earth phase is compensated at the same time [4].

Generally, BP consists of 3 operators, which can be denoted as:

$$\mathbf{X}_m = \operatorname{unfold}\left(\mathbf{H}_{mA} \operatorname{vec}\left(\hat{\mathbf{F}}_R^H \mathbf{P}_D\left(\mathbf{H}_R \odot (\mathbf{F}_R \mathbf{Y}_m)\right)\right)\right)$$
$$\mathbf{X}_s = \operatorname{unfold}\left(\mathbf{H}_{sA} \operatorname{vec}\left(\hat{\mathbf{F}}_R^H \mathbf{P}_D\left(\mathbf{H}_R \odot (\mathbf{F}_R \mathbf{Y}_s)\right)\right)\right) \quad (7)$$

where $\mathbf{Y}_m$ and $\mathbf{Y}_s$ are the received echoes of the master and the slave antenna respectively. And $\mathbf{X}_m = \mathbf{X} \odot \mathbf{P}_m$ and $\mathbf{X}_s = \mathbf{X} \odot \mathbf{P}_s$ are the imaging results.

The first operator is MF in the range direction, which can be implemented fast in the frequency domain, corresponding to $\mathbf{H}_R \odot (\mathbf{F}_R \mathbf{Y}_m)$ and $\mathbf{H}_R \odot (\mathbf{F}_R \mathbf{Y}_s)$ in (7). $\mathbf{F}_R$ is the DFT matrix. $\mathbf{H}_R$ is the match-filtering matrix [3].

The second operator is interpolation in the range direction. It can be implemented through zero-padding in the frequency domain, corresponding to $\hat{\mathbf{F}}_R^H \mathbf{P}_D(\cdot)$. $\hat{\mathbf{F}}_R^H$ is the and inverse DFT matrix and $\mathbf{P}_D = [\mathbf{I} \ \mathbf{0} \ \mathbf{I}]^T$ is the padding matrix, where $\mathbf{I}$ is the truncated unit matrix.

The third operator is to compensate Doppler phase and to accumulate in the azimuth direction, corresponding to $\operatorname{unfold}(\mathbf{H}_{mA}\operatorname{vec}(\cdot))$ and $\operatorname{unfold}(\mathbf{H}_{sA}\operatorname{vec}(\cdot))$. And operators of and its adjoint are respectively indicated as $\operatorname{vec}(\cdot)$ and $\operatorname{unfold}(\cdot)$. $\mathbf{H}_{mA}$ and $\mathbf{H}_{sA}$ are corresponding matrices of phase compensation and accumulation, which are expressed as:

$$\mathbf{H}_{mA} = \begin{bmatrix} \mathbf{h}_{m1}^H & \cdots & \mathbf{h}_{mN}^H \end{bmatrix}^H \quad \mathbf{H}_{sA} = \begin{bmatrix} \mathbf{h}_{s1}^H & \cdots & \mathbf{h}_{sN}^H \end{bmatrix}^H \quad (8)$$

They are stacks of compensation and accumulation vectors for all the imaging pixels. For a single imaging cell, taking the master antenna for example, it can be expressed as:

$$\mathbf{h}_{mi} = \begin{bmatrix} \cdots & \mathbf{0} & \exp(j4\pi R_{mi}(n)/c) & \mathbf{0} & \cdots \end{bmatrix} \quad (9)$$

This vector is the composition of phase compensation terms $\exp(j4\pi R_{mi}(n)/c)$ of all azimuth time, where $R_{mi}(n)$ is the instant slant range and $c$ indicates the light speed. It is noted that its position is varied at different azimuth time, which can be calculated as:

$$\mathrm{id}\left(\exp\left(j4\pi R_{mi}(n)/c\right)\right) = \lfloor 2NF_s\left(R_{mi}(n) - R_0\right)/c \rfloor \quad (10)$$

Where $F_s$ is the sampling frequency and $R_0$ is the reference slant range. $\lfloor \cdot \rfloor$ is the rounding operator.

The interferogram can be obtained as:

$$\mathbf{\Theta} = \mathbf{X}_m \odot \mathrm{conj}(\mathbf{X}_s) \quad (11)$$

Thus, the BP embedded forward observation process can be expressed as:

$$\mathbf{Y}_m = \mathbf{F}_R^H\left(\mathbf{H}_R^* \odot \left(\mathbf{P}_D^H \hat{\mathbf{F}}_R \mathrm{unfold}\left(\mathbf{H}_{mA}^H\left(\mathrm{vec}(\mathbf{X}_S \odot \mathbf{\Theta})\right)\right)\right)\right) + \mathbf{N} \quad (12)$$

where $\mathbf{H}_R^* = conj(\mathbf{H}_R)$. Its corresponding form of matrix-vector can be expressed as:

$$\mathbf{y}_m = \mathbf{A}\mathbf{\theta} + \mathbf{n} \quad (13)$$

where

$$\mathbf{A} = \left(\mathbf{I} \otimes \mathbf{F}_R^H\right)\mathrm{diag}\left(\mathbf{H}_R^*\right)\left(\mathbf{I} \otimes \left(\mathbf{P}_D^H \hat{\mathbf{F}}_R\right)\right)\mathbf{H}_{mA}^H \mathrm{diag}(\mathbf{X}_S) \quad (14)$$

In (14), $\otimes$ indicates the Kronecker product, and $\mathrm{diag}(\cdot)$ denotes the diagonal operator [1]. $\mathbf{y}_m$, $\mathbf{\theta}$ and $\mathbf{n}$ are vectorizations of $\mathbf{Y}_m$, $\mathbf{\Theta}$ and $\mathbf{N}$ respectively. (14) is the composition of adjoint operators of BP.

## 3. INTERFEROMETRIC PHASE FORMATION

In combination with (6), (13) and (14), the interferogram is solved through the following optimization problem:

$$\arg\min_{\mathbf{\theta}} \frac{1}{2}\|\mathbf{y}_m - \mathbf{A}\mathbf{\theta}\|_2^2 + \lambda\|\Psi(\mathbf{\Theta})\|_1 \quad (15)$$

It can be solved forward-backward iteration by proximal gradient descent algorithm [5]. Specifically, we take the $\Psi$ as the Fourier transform for example, then (16) can also be expressed as:

$$\arg\min_{\mathbf{\theta}_f} \frac{1}{2}\|\mathbf{y}_m - \mathbf{A}\mathbf{F}_{AR}\mathbf{\theta}_f\|_2^2 + \lambda\|\mathbf{\theta}_f\|_1 \quad (16)$$

where $\mathbf{F}_{AR}$ is the inverse 2D Fourier transform matrix. Seen from (14), the forward observation process contains the second antenna's imaging result, which can't be directly obtained. Here we use the result of BP to approximate. At the iteration $k$, two steps are performed as follows:

1. Take a forward-backward step to fulfill gradient descent.

$$\mathbf{z} = \mathbf{\theta}_f^{k-1} + \mu \mathbf{F}_{AR}^H \mathbf{A}^H\left(\mathbf{y}_m - \mathbf{A}\mathbf{F}_{AR}\mathbf{\theta}_f^{k-1}\right) \quad (17)$$

where $\mu$ is the iteration step of gradient descent. $\mathbf{AF}_{AR}\mathbf{\theta}_f^{k-1}$ is a forward process, which can be considered as the echo generation, which can be modified as:

$$R^{k-1} = \mathbf{F}_R^H\left(\mathbf{H}_R^* \odot \left(\mathbf{P}_D^H \hat{\mathbf{F}}_R \mathbf{Y}_{ma}^{k-1}\right)\right) \quad (18)$$

where

$$\mathbf{Y}_{ma}^{k-1} = \mathrm{unfold}\left(\mathbf{H}_{mA}^H \mathrm{vec}\left(\mathbf{X}_S \odot \left(\mathbf{F}_R^H \mathbf{\Theta}_f^{k-1}\mathbf{F}_A\right)\right)\right) \quad (19)$$

$\mathbf{Y}_{ma}^{k-1}$ can be seen as the interpolated range-compressed echo of the scene $\mathbf{X}_S \odot \left(\mathbf{F}_R^H \mathbf{\Theta}_f^{k-1}\mathbf{F}_A\right)$. Practically, the Fourier Transform can be computed by the fast Fourier transform (FFT). And the echo is calculated point-wise, which can be implemented fast in parallel. Further, the backward process is $\mathbf{F}_{AR}^H \mathbf{A}^H(\cdot)$ that is similar to BP, which can also be put in parallel.

2. Perform element-wise soft-thresholding:

$$\mathbf{\theta}_f^k = S_{\mu\lambda}(\mathbf{z}) \quad (20)$$

where $\left[S_{\mu\lambda}(\mathbf{z})\right]_j = \mathrm{sign}(z_j)\max\left(|z_j| - \mu\lambda, 0\right)$. $\mathrm{sign}(\cdot)$ is the complex sign function.

The inevitable speckle effect exists in the SAR image, which causes that the phase of $\mathbf{X}_s$ is random. Thus, randomness is induced into the forward observation $\mathbf{A}$, which ensures the reconstruction successfully [1]. In our experiments, normally, satisfying results can be obtained within 5 iterations.

## 4. EXPERIMENT

We test the proposed method based on both simulated and measured data. For simulation, the scene containing a cone is considered, shown in Fig. 2. The maximum height is 40m, covering an area of $300\,\mathrm{m} \times 300\,\mathrm{m}$. And the InSAR system flights at an elevation of $3000\,\mathrm{m}$ with a speed of $50\,\mathrm{m/s}$. The baseline length is $1\,\mathrm{m}$ and the incident angle is $35°$. The signal bandwidth is 400 MHz centering at 35 GHz.

Results are shown in Fig. 3. The ideal interferometric phase is shown at the upper right in Fig. 2. The results of the conventional method are shown in Fig. 3(a)-(b). BP is chosen to compare, specifically. The results of the proposed method are shown in Fig. 3(c)-(d). Two sampling situations are tested, 100% full-sampling and 50% random-sampling. Visually, for both the two sampling situations, the proposed method obtains both higher-quality interferogram with rarely noise existence. Further, we calculate the root mean square error (RMSE) [7], shown in Fig. 3. Compared with BP, the results of our method are **less in both sampling situations**.

We further test the proposed method on the measured data from a Ka-band air-borne InSAR system. The length of baseline length is 0.3m and the incident angle is $49.6°$. The imaging scene is a slope area in the suburbs, shown in Fig. 4. The curved area across the middle is a river, while area of farmland with slow-varying terrain is setting on the two sides. The result of BP in combination with the filtering by InSAR-BM3D [7] is chosen as reference, shown in Fig. 4. Results are shown in Fig. 5. Compared with the results of BP, the proposed method obtains higher phase quality with less noise and more local details. These experiments verify the effectiveness of the proposed method.

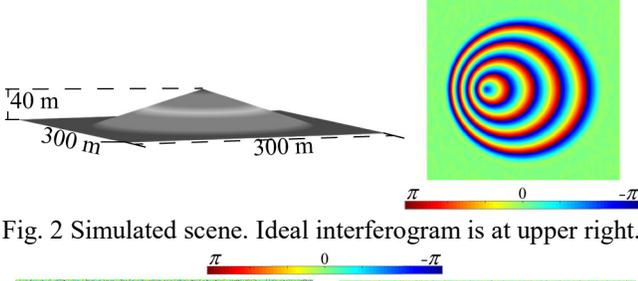

Fig. 2 Simulated scene. Ideal interferogram is at upper right.

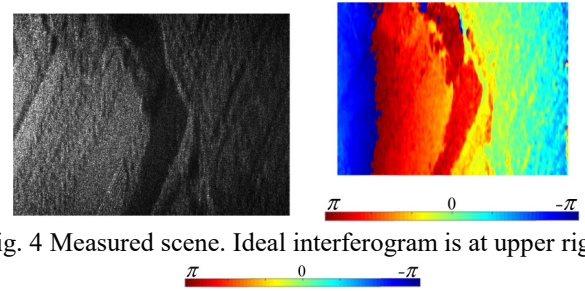

Fig. 4 Measured scene. Ideal interferogram is at upper right.

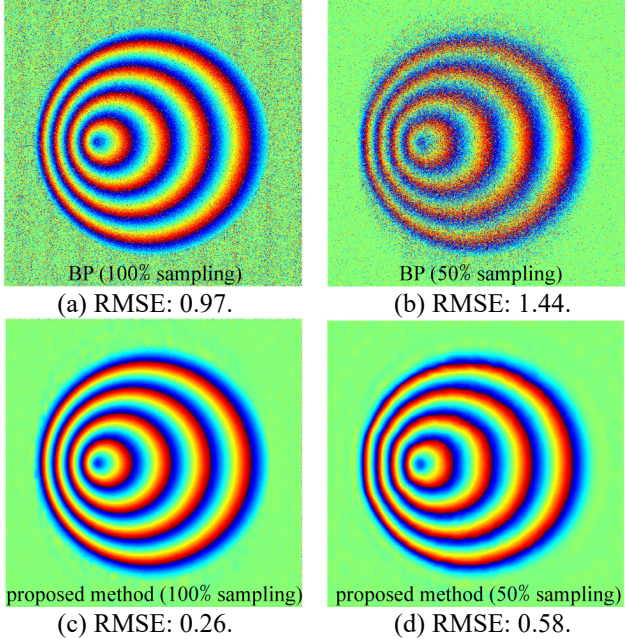

Fig. 3 Simulation Results. (a)-(b) results of BP; (c)-(d) results of the proposed method;

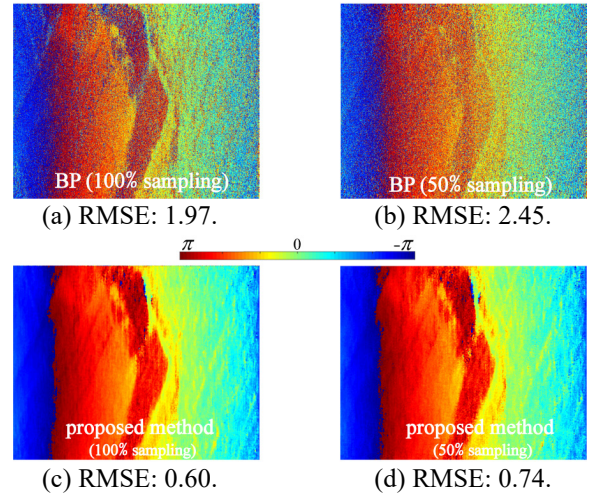

Fig. 5 Measured data Results. (a)-(b) results of BP; (c)-(d) results of the proposed method;

## 5. CONCLUSION

In this paper, we propose a new InSAR imaging method for obtaining higher quality interferogram directly from echoes, without post-processing. We construct the imaging model based on 2D sparse regularization to embed characteristics of scene. Besides, BP is also embedded to approximate the forward process. Then, the forward-backward iteration is formed to solve the model that can implemented in parallel inheriting merits of BP. Experiments verify the effectiveness of our method. Compared with the conventional method, it shows the superiority in terms of imaging quality and processing complexity.